%% file: algo.tex
\documentclass[12pt]{article}
\usepackage{amssymb}
\usepackage{epsfig}
\usepackage{sint,cite}

\newcommand\be{\begin{equation}}
\newcommand\ee{\end{equation}}
\newcommand\bea{\begin{eqnarray}}
\newcommand\eea{\end{eqnarray}}

\newcommand{\rmO}{{\rm O}}

\def\rmP{{\rm P}}
\def\ZP{Z_{\rmP}}

\newcommand{\fP}{f_{\rm P}}
\newcommand{\fA}{f_{\rm A}}
\newcommand{\cA}{c_{\rm A}}

\def\npf{n_{\rm pf}}
\def\Pa{P_{\rm acc}}
\def\DH{\Delta H}
\def\DHc{|\DH({\rm cycle})|}
\def\DU{||\Delta U||}
\def\Iam1{\langle\exp(-\DH)\rangle}
\def\eps{\varepsilon}
\def\tauint{\tau_{\rm int}}

\def\gbar{\bar{g}}
\def\mbar{\overline{m}}
\def\SUthree{{\rm SU(3)}}

\def\pp{\prime\prime}
\newcommand{\eq}[1]{eq.~(\ref{#1})}
\newcommand{\fig}[1]{Fig.~\ref{#1}}
\newcommand{\tab}[1]{Table~\ref{#1}}
\newcommand{\sect}[1]{Section~\ref{#1}}
\def\Dw{D_{\rm w}}
\def\fm{\,{\rm fm}}
\def\MD{{\rm MD}}
\def\sec{{\rm sec}}

\newcommand{\ev}[1]{\langle #1 \rangle}

\def\ma[#1,#2,#3,#4]  {{\left( \matrix{ #1  & #2 \cr
                                        #3  & #4 \cr } \right)}}

\begin{document}

\thispagestyle{empty}
\input{title}

\input{sect1.tex}
\input{sect2.tex}
\input{sect3.tex}
\input{sect4.tex}

\begin{appendix}
\input{appa.tex}
\end{appendix}

\bibliography{algo}           
\bibliographystyle{h-elsevier}   

\end{document}

%% file: title.tex
\title{{\normalsize\vskip -50pt
\mbox{} \hfill DESY 03-071 \\
\mbox{} \hfill HU-EP-03/34 \\
\mbox{} \hfill SFB/CPP-03-11 \\}
\vskip 25pt
Simulating the Schr{\"o}dinger functional with two pseudo-fermions}

\author{
\centerline{
            \epsfxsize=2.5 true cm
            \epsfbox{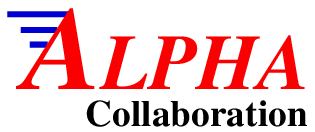}}\\
Michele Della Morte $^{a}$, Francesco Knechtli $^{b}$, Juri Rolf, $^{b}$\\
Rainer Sommer $^{a}$, Ines Wetzorke $^{c}$, Ulli Wolff $^{b}$\\[1cm]
$^{a}$ DESY, Platanenallee 6, 15738 Zeuthen, Germany\\[0.5cm]
$^{b}$ Institut f\"ur Physik, Humboldt Universit\"at,\\
Newtonstr. 15, 12489 Berlin, Germany\\[0.5cm]
$^{c}$ NIC/DESY-Zeuthen,\\ Platanenallee 6, 15738 Zeuthen, Germany\\[1cm]
}

\maketitle

\begin{abstract}
  We report on simulations with two flavors of
  O($a$) improved degenerate Wilson fermions with
  Schr{\"o}dinger functional boundary conditions.
  The algorithm which is used is Hybrid Monte Carlo
  with two pseudo-fermion fields as proposed by M. Hasenbusch.
  We investigate the numerical precision and sensitivity to
  reversibility violations of this algorithm.
  A gain of a factor two in CPU cost is reached compared with
  one pseudo-fermion field due to the larger possible step-size.
\end{abstract}


\newpage

%% file: sect1.tex
\section{Introduction \label{intro}}

Algorithmic tests with dynamical fermions are large scale projects by themselves.
It is therefore a good idea to integrate them in large scale physics projects,
part of whose runs are dedicated to study variants of the algorithm or
whose configurations are stored and used to further analyze the algorithm.
In this work we present a study of a variant of the HMC algorithm which uses
two pseudo-fermion fields per degenerate flavor doublet,
as proposed by M. Hasenbusch \cite{Hasenbusch:2001ne,Hasenbusch:2002ai}
and recently tested in \cite{AliKhan:2003br}. Our simulations
of full QCD are performed within the Schr{\"o}dinger functional framework
and deal with two flavors of massless quarks. The Wilson plaquette action
is taken for the gauge field and the $\rmO(a)$ improved Wilson-Dirac operator
for the fermions.

The algorithmic study presented here is integrated in the ALPHA project for
the computation of the running of the renormalized quark mass
\cite{Knechtli:2002vp}. Since we use a finite size technique our lattices
range from a very small physical size of $10^{-2}\fm$ to an intermediate
size of $1\fm$. For these lattice volumes we can study the scaling behavior
of the HMC algorithm since we can simulate different lattice spacings
at fixed physical lattice size.

In \sect{twopf} we describe the main steps to build the HMC algorithm with
two pseudo-fermion fields. We explain our choice of the factorization
of the Dirac operator into two parts. \sect{numres}
is dedicated to the study of the required numerical precision and to the
reversibility of the integration of the molecular dynamics equations of motion.
A comparison with HMC using one pseudo-fermion field is done for several physical
lattice sizes and shows that we gain about a factor two in performance when
using two pseudo-fermion fields.

%% file: sect2.tex
\section{HMC with two pseudo-fermion fields \label{twopf}}

We would like to simulate on the lattice two flavors of Wilson quarks
in the framework of the Schr{\"o}dinger functional with $\rmO(a)$ improvement
\cite{Luscher:1996sc}.
The Hamiltonian for the evolution in fictitious time is defined as
\begin{equation}\label{Hamiltonian}
 H = \frac{1}{2}\sum_{x,\mu}{\rm Tr}[\pi(x,\mu)^2] + S_{\rm eff} \,,
\end{equation}
where $\pi(x,\mu)$ are the traceless Hermitian momenta conjugate
to the gauge field link variables $U(x,\mu)$ and the effective action
(at first for one pseudo-fermion field) is given by
\begin{equation}\label{s_eff1}
 S_{\rm eff} = S_{\rm g}(U) + S_{\rm det}(U) + 
 S_{\rm PF}(U,\phi,\phi^{\dagger}) \,.
\end{equation}
In \eq{s_eff1} $S_g(U)$ is the Wilson plaquette gauge action 
with gauge coupling $g_0^2=6/\beta$
and the other two terms represent the fermionic contributions. 
The Wilson-Dirac operator is
\be
 a(\Dw + \delta D + m_0) = \frac{1}{2\kappa}M \,, 
 \quad \kappa=(8+2am_0)^{-1} \,,
\ee
where $\delta D$ is the correction for $\rmO(a)$ improvement (clover term).
We use even-odd preconditioning \cite{DeGrand:1990dk} and the Wilson-Dirac
operator assumes the block form
\be
 M = \left( 
 \begin{array}{cc} 1+T_{ee} & M_{eo} \\ M_{oe} & 1+T_{oo} \end{array}
 \right) \,,
\ee
where the subscripts $e$ ($o$) refer to the even (odd) sites of the lattice.
The boundary values of the quark fields are set to zero and in this case
the $\rmO(a)$ improvement terms proportional to $(\tilde{c}_t-1)$ 
do not contribute to the off-diagonal blocks \cite{Luscher:1996vw}.
The operators $T_{ee}$ and $T_{oo}$ are diagonal in space-time and represent
the $\rmO(a)$ corrections.
The Hermitian operator $\gamma_5M$ can be written as the product
\be
 \left(
 \begin{array}{cc} \gamma_5(1+T_{ee}) & 0 \\ \gamma_5M_{oe} & 1 \end{array}
 \right) \times \left(
 \begin{array}{cc} 1 & (1+T_{ee})^{-1}M_{eo} \\ 0 & 
 \gamma_5\{1+T_{oo}-M_{oe}(1+T_{ee})^{-1}M_{eo}\} 
 \end{array}\right) \,.
\ee
We define the following Hermitian operators
\begin{eqnarray}
& Q=c_0\gamma_5M\,, & c_0=(1+8\kappa)^{-1} \,, \label{Q} \\
& \hat{Q}=\tilde{c}_0\gamma_5\{1+T_{oo}-M_{oe}(1+T_{ee})^{-1}M_{eo}\}\,, &
\tilde{c}_0=(1+64\kappa^2)^{-1} \,. \label{Qhat}
\end{eqnarray}
The fermionic contribution to the partition function can then be written as
\begin{equation}\label{fermionicZ}
\det(Q^2) \propto \det(1+T_{ee})^2\det\hat{Q}^2 \,.
\end{equation}
From \eq{fermionicZ} we get the contributions to $S_{\rm eff}$ in
\eq{s_eff1}
\begin{eqnarray}
S_{\rm det} & = & -2{\rm tr}\ln(1+T_{ee}) \label{s_det} \\
\det(\hat{Q}^2) & \propto & \int{\rm D}[\phi]\,\exp(-S_{\rm PF}) \,,\quad
S_{\rm PF} = \phi^{\dagger}\hat{Q}^{-2}\phi \,, \label{s_pf}
\end{eqnarray}
in terms of one dynamical complex pseudo-fermion field $\phi$.

In \cite{Hasenbusch:2001ne,Hasenbusch:2002ai} a modified effective pseudo-fermionic
action has been proposed.
By using the identity $\hat{Q}=\tilde{Q}\tilde{Q}^{-1}\hat{Q}$ for some arbitrary
invertible matrix $\tilde{Q}$, we can compute $\det(\hat{Q}^2)$ by using two
instead of one pseudo-fermion fields
\begin{eqnarray}
\det(\hat{Q}^2) & \propto & \int\left(\prod_{i=1}^{\npf}{\rm D}[\phi_i]\right)\,
\exp(-\sum_{i=1}^{\npf}S_{{\rm F}_i}) \,, \quad \npf=2 \\
S_{{\rm F}_1} & = & \phi_1^{\dagger}(\tilde{Q}\tilde{Q}^{\dagger})^{-1}\phi_1
\label{s_pf1} \\
S_{{\rm F}_2} & = & \phi_2^{\dagger}
(\hat{Q}^{-1}\tilde{Q}\tilde{Q}^{\dagger}\hat{Q}^{-1})\phi_2 \,. \label{s_pf2}
\end{eqnarray}
In this work we choose
\begin{equation}\label{rhoshift}
\tilde{Q} = \hat{Q}-i\rho \,,
\end{equation}
where $\rho$ is a real number. We can then cast \eq{s_pf1} and \eq{s_pf2}
into the form
\begin{equation}\label{s_pfsplit}
S_{{\rm F}_i} = \phi_i^{\dagger}\left[\sigma_i^2+(\hat{Q}^2+\rho_i^2)^{-1}\right]
\phi_i \,, \quad i=1,2 \,,
\end{equation}
where the field $\phi_2$ has been rescaled and
\begin{eqnarray}
\sigma_1=0\,, && \rho_1=\rho\,, \label{rspar1} \\
\sigma_2=\frac{1}{\rho}\,, && \rho_2=0\,. \label{rspar2}
\end{eqnarray} 
In the Hamiltonian \eq{Hamiltonian} the effective action is now given by
\begin{equation}\label{s_eff2}
 S_{\rm PF} = 
 \sum_{i=1}^2S_{{\rm F}_i}(U,\phi_i,\phi_i^{\dagger}) \,.
\end{equation}
We determine the parameter $\rho$ in \eq{rhoshift} by requiring that the
sum $K$ of the condition numbers of the operators appearing in the actions
$S_{{\rm F}_1}$ and $S_{{\rm F}_2}$ is minimal.
If we use the notation $\lambda_{\min}\equiv\lambda_{\min}(\hat{Q}^2)$, 
$\lambda_{\max}\equiv\lambda_{\max}(\hat{Q}^2)$ for the lowest respectively highest
eigenvalue of $\hat{Q}^2$ we get
\begin{eqnarray}
 K & = & y(\rho) + \frac{k}{y(\rho)} \,,\quad
 y(\rho) = \frac{\lambda_{\max}+\rho^2}{\lambda_{\min}+\rho^2} \,,
\end{eqnarray}
where
\be\label{cn}
  k = \frac{\lambda_{\max}}{\lambda_{\min}}
\ee
is the condition number of $\hat{Q}^2$.
The minimal $K$ is reached for $y(\rho)=\sqrt{k}$.
The condition numbers of the operators in $S_{{\rm F}_1}$ and $S_{{\rm F}_2}$
are then both equal to $\sqrt{k}$. Solving for $\rho$ we obtain
$\rho = (\lambda_{\min}\lambda_{\max})^{1/4}$. We set\footnote{
In practice $\lambda_{\max}$ fluctuates negligibly compared to $\lambda_{\min}$.}
\begin{equation}\label{rhochoice}
 \rho = \left(\ev{\lambda_{\min}}\ev{\lambda_{\max}}\right)^{1/4} \,,
\end{equation}
where we denote by $\ev{\cdot}$ the expectation value computed
in a Monte Carlo simulation. Of practical advantage is the
following relation which holds at tree level in the Schr{\"o}dinger functional
\cite{Sint:1994un}
\bea
 k & \propto & \left(\frac{T}{a}\right)^2 \,, \label{ksf}
\eea
where $T$ is the temporal extension of the box.
Qualitatively the same behavior (at the critical mass)
is expected for
\bea
 \ev{k} = \Big\langle\frac{\lambda_{\max}}{\lambda_{\min}}\Big\rangle 
 & \propto & \left(\frac{T}{a}\right)^2 \label{cnev}
\eea
in the continuum limit at fixed
renormalized coupling $\gbar^2(L)=u$, where $L$ is the spatial extension of the box.
This means that $\rho$ as defined in \eq{rhochoice}
should be scaled with the lattice size $T/a$ approximately like $\sqrt{a/T}$.

The algorithm, which we use in our simulations of \eq{Hamiltonian} and
\eq{s_eff2} is Hybrid Monte Carlo
\cite{Duane:1987de,Gottlieb:1987mq} and consists
of the following 4 steps:\\[1ex]
\noindent{\bf 1.}
The momenta $\pi(x,\mu)$ are generated from the Gaussian distribution of
their components in a basis of the Lie Algebra of $\SUthree$.\\[1ex]
\noindent{\bf 2.}
The two pseudo-fermion fields $\phi_1$ and $\phi_2$ are generated by global
heatbath according to the probability distribution
\begin{equation}\label{probphi}
P(\phi_i) \propto \exp\left\{-\phi_i^{\dagger}
\left[\sigma_i^2+(\hat{Q}^2+\rho_i^2)^{-1}\right]\phi_i\right\}
\propto \exp\{-R^{\dagger}R\} \,.
\end{equation}
This is done as usual by generating a complex Gaussian random vector $R$.
For $\sigma_i\neq0$ we use
\begin{equation}
(\hat{Q}^2+\rho_i^2+\sigma_i^{-2})\phi_i =
\sigma_i^{-1}(\hat{Q}-i\sqrt{\rho_i^2+\sigma_i^{-2}})(\hat{Q}-i\rho_i)R \,.
\end{equation}
If $\sigma_i=0$ we use
\begin{equation}
\phi_i = (\hat{Q}-i\rho_i)R \,.
\end{equation}
\noindent{\bf 3.}
The gauge links $U(x,\mu)$ and the momenta $\pi(x,\mu)$ are evolved along
a trajectory of length $\tau$ by integrating the
molecular dynamics equations of motion with step-size $\delta\tau$.
This can be done in full analogy with \cite{Jansen:1997yt}.
In the equations of motion for the momenta we need the variation
$\delta S_{{\rm F}_i}$ of the pseudo-fermion actions under an infinitesimal
change of the gauge link $\delta U(x,\mu)$.
With the help of the vectors
\begin{eqnarray}
 X_i & = & (\hat{Q}^2+\rho_i^2)^{-1}\phi_i \,, \label{Xvec} \\
 Y_i & = & (\hat{Q}+i\rho_i)X_i \,, \label{Yvec}
\end{eqnarray}
defined on the odd sites of the lattice we construct over the full lattice
the vectors
\be\label{xbarybar}
 \overline{X}_i=\left( \begin{array}{c} -(1+\delta M_{ee})^{-1}M_{eo}X_i \\ X_i
 \end{array} \right) \; , \;
 \overline{Y}_i=\left( \begin{array}{c} -(1+\delta M_{ee})^{-1}M_{eo}Y_i \\ Y_i
 \end{array} \right) \,,
\ee
which we use to write
\begin{equation}\label{forceQ}
 \delta S_{{\rm F}_i} = -\frac{\tilde{c}_0}{c_0}(
 \overline{Y}_i^{\dagger}\delta Q\overline{X}_i+
 \overline{X}_i^{\dagger}\delta Q\overline{Y}_i) \,.
\end{equation}
The variation of the two pseudo-fermion actions can be expressed in terms of
the variation of the operator $\delta Q$, in the same way as for the
standard version with one pseudo-fermion field.

As integration scheme we use an improved
Sexton-Weingarten \cite{Sexton:1992nu} integrator with different
time scales for the gauge part and for the pseudo-fermion part of the force
which governs the evolution of the momenta. Our integration
scheme is the same as the one used in \cite{Jansen:1995gz} and partially removes
the step-size errors $\rmO((\delta\tau)^3)$ in one integration step.\\[1ex]
\noindent{\bf 4.}
Metropolis accept/reject step: the new configuration
$\{\pi^{\prime},U^{\prime}\}$ is accepted with probability
\be
 \min\left\{1,\exp(-\DH)\right\} \,,
\ee
where
\begin{equation}\label{dham}
 \DH = H(\pi^{\prime},U^{\prime},\phi_i) - H(\pi,U,\phi_i) \,.
\end{equation}
We monitor $\exp(-\DH)$ to check for the correctness of the algorithm,
in particular this quantity has proved to be sensitive to the required numerical
reversibility in the integration
of the equations of motion. 
For Metropolis-like algorithms it is possible to prove the general property
\bea
 \Iam1 & = & 1 \,. \label{Iam1=1}
\eea
We emphasize that this expectation value is sensitive to {\em all}
proposed field configurations not only to the accepted ones.

%% file: sect3.tex
\section{Numerical results \label{numres}}

We present a numerical study of simulations of two degenerate massless quarks
in the $\rmO(a)$ on--shell improved Schr{\"o}dinger functional using the algorithm
presented in the previous section. For the lattice temporal and spatial extensions,
the background gauge field and the parameter $\theta$ controlling the spatial
boundary conditions of the fermion fields we set respectively
\be\label{SFparams}
  T=L\,, \quad C=C^{\prime}=0\,, \quad \theta=0.5 \,.
\ee
(see \cite{Luscher:1996sc} for more detailed definitions of these quantities). 
This choice is motivated by the fact that our algorithmic study is part of
large scale simulations performed by the ALPHA collaboration to determine the
running of the renormalized quark mass in two flavor QCD \cite{Knechtli:2002vp}.
The massless theory is defined in the bare parameter space along the line
$\kappa=\kappa_c(\beta)$ where the PCAC mass
\be\label{PCACmass}
 m = \left. \frac{\frac{1}{2}(\partial_0^{\ast}+\partial_0)\fA(x_0)
                +\cA a\partial_0^{\ast}\partial_0\fP(x_0)}
               {2\fP(x_0)}\right|_{x_0=T/2}
\ee
vanishes\footnote{
Note that many choices of matrix elements of the PCAC relation yield
masses which agree up to $\rmO(a^2)$ effects. The mass definitions can differ
for instance in the choice of the parameters \eq{SFparams}, of $x_0$
\eq{PCACmass} or of the boundary states.}.
Here, $\partial_0$ and $\partial_0^{\ast}$ are the forward and backward lattice
derivatives, respectively, and $\cA$ denotes the coefficient multiplying the
O($a$) improvement term in the improved axial current. For the definitions of the
correlation functions $\fA$ and $\fP$ we follow \cite{Capitani:1998mq}.
We stress the fact that simulations along the massless line are possible
since the Schr{\"o}dinger functional has a natural infrared cut-off proportional
to $1/T^2$ in the spectrum of the Dirac operator squared \cite{Sint:1994un}.
If the fluctuations of the gauge field are not too large this cut-off avoids
the occurrence of small eigenvalues and thus makes simulations of the
massless theory possible in not too large physical volumes.

In the Schr{\"o}dinger functional
the renormalized coupling $u=\gbar^2(L)$ is presumably a monotonically
growing function of $L$ \cite{Bode:2001jv,Heitger:2001hs,DellaMorte:2002vm}. Keeping
the physical size $L$ constant is equivalent to holding the renormalized coupling
fixed. We perform simulations approximately at couplings
\be
 u = 1.0\,,\; 1.1\,,\; 1.2\,,\; 1.3\,,\;
     1.5\,,\; 1.8\,,\; 2.0\,,\; 2.5\,,\; 3.3\,,\; 5.7 \,,
 \label{uvalues}
\ee
and for each coupling at various lattice sizes $L/a$ corresponding to different
lattice resolutions. A summary of the bare parameters and the formulae 
used for
the O($a$) improvement coefficients as functions of the bare gauge coupling
can be found in Appendix \ref{simpar}.
Moreover we did some exploratory simulations at $\beta=5.2$ and
$\kappa=0.1355$ with lattice sizes $L/a$=6,\,8 and 12. The corresponding
renormalized coupling is presently only known for $L/a=6$ to be $u = 4.7$.
The value $\kappa=0.1355$ is the same as for the lightest quark mass used
in the large volume simulations of Refs. \cite{Allton:2001sk,Aoki:2002uc}.
As long as a low energy reference scale is not determined,
we cannot yet associate the $u$ values with physical units of the box size $L$.
We use the renormalized couplings to ``label'' our simulations. Indicatively
the couplings in \eq{uvalues} correspond to the range $10^{-2}\fm\ldots1\fm$.

In the simulations we report on here the quantity we are primarily interested in
is the renormalization constant of the pseudoscalar density $\ZP$. For a definition
of $\ZP$ see \cite{Capitani:1998mq}. In the massless
renormalization scheme that we employ, $\ZP$ depends on the size
of the system and on the lattice resolution.
We are interested in $\ZP$ since its running with the renormalization scale
$\mu=1/L$ determines the running of the renormalized quark mass
$\mbar(\mu)$:
\bea
 \frac{\mbar(\mu)}{\mbar(\mu/2)} & = & \left.
 \lim_{a/L\to0}\,\frac{\ZP(g_0,2L/a)}{\ZP(g_0,L/a)}\right|_{
 \gbar^2(L)=u} \,.
\eea
As a measure of the CPU cost of our simulations we take the integrated
autocorrelation time of $\ZP$ per lattice point
\begin{equation}\label{tauint}
 \tauint(\ZP)\;[\sec] = \tauint(\ZP)\;[\MD]\times t_{\MD}\times
 \left(\frac{a}{T}\right)\left(\frac{a}{L}\right)^3 \,.
\end{equation}
We denote by $\tauint(\ZP)\;[\MD]$ the integrated autocorrelation time
\cite{Wolff:2003sm} in units of molecular dynamics (MD) time.
In \eq{tauint} $t_{\MD}$ are the
CPU seconds needed on one APEmille crate
for one update move of one unit of MD time.
Besides being machine-dependent
$t_{\MD}$ depends on the algorithm and on the step-size $\delta\tau$.
One APEmille crate consists of 128 nodes whose peak performance adds up to 68 GFlops.
A trivial volume factor is divided out in \eq{tauint}.
One call of the operator $\hat{Q}$ on APEmille costs roughly
$12\,\mu\,{\rm sec}$ per lattice point (depending on communication).
By dividing the value of $\tauint(\ZP)\;[\sec]$ by
$12\,\mu\,{\rm sec}$ we get the integrated autocorrelation time expressed in 
the equivalent number of applications of $\hat{Q}$.

\subsection{Numerical precision \label{subpre}}

We carry out the numerical simulations in single precision arithmetics\footnote{
Global sums are performed in double precision.}.
In this section we study effects of the numerical precision of our simulations using
$\npf=2$ pseudo-fermion fields by changing some of the algorithmic parameters.

During the integration of the molecular dynamics equations of motion
the inversions of the operators $\hat{Q}^2+\rho^2$ and $\hat{Q}^2$ required
in \eq{Xvec} are performed
with the conjugate gradient (CG) iteration algorithm. We use as a stopping criterion
the relative residue which is defined as
the square of the ratio between the norm of the residue vector and the norm of
the present solution vector. The iteration stops whenever this number reaches
the requested accuracy $\eps^2$. The starting vector for each inversion along
the trajectory is chosen to be zero.
We first investigated the possibility of increasing the relative residue compared
to the single precision value $\eps^2=10^{-13}$, thereby saving CPU time.
We emphasize that detailed balance holds for all values of $\eps^2$.
In \tab{t_eps} we compare the results for the choice $\eps^2=10^{-10}$.
We observe a 2$\sigma$ deviation in $\ZP$ and an almost 5$\sigma$ deviation
of $\Iam1$ from $1$.
 \begin{table}[htb] 
   \centering
   \begin{tabular}{|ccccccc|}
    \hline
    $\tau$ & $\eps^2$ & $\ZP$ & $\tauint(\ZP)\;[\MD]$ & plaq. & $am$ & $\Iam1$ \\[0.5ex]
  \hline\hline
1 & $10^{-13}$ & 0.6236(17) & 22(8) & 0.68756(1) & $-0.00122(3)$ & 0.989(7) \\
1 & $10^{-10}$ & 0.6288(22) & 32(12) & 0.68756(1) & $-0.00118(3)$ & 1.038(8) \\
  \hline 
 \end{tabular} 
\caption{Comparison of simulation results for $\npf=2$ with the relative residue
set to the full single precision value $\eps^2=10^{-13}$ and to $\eps^2=10^{-10}$.
Results from $L/a=24$ lattices at coupling 2.5.}
\label{t_eps}
\end{table}

We have recently learned that
our definition of the relative residue $\eps^2$ unfortunately
differs from the one commonly used
in the numerical mathematics literature. There $\eps^2$ is defined as the square of
the ratio between the norm of the residue vector and the norm of the source vector,
which is independent of the normalization of the operator inverted on
the source vector. We checked the difference between these two definitions of
$\eps^2$ in the simulations of \tab{t_eps}. When $\eps^2$ normalized using the
solution vector reaches the requested precision $10^{-13}$ or $10^{-10}$, the
values of $\eps^2$ normalized using the source vector are about 10 times larger.
To reach $\eps^2=10^{-13}$ defined with the latter normalization it requires an
increase of 25\% in CG iterations.

Relaxing the relative residue in the CG inversion leads to a systematic violation 
of \eq{Iam1=1}. Since this possibly
indicates the violation of reversibility in the integration of the equations of
motion we investigate this issue in more detail in the next section.

\subsection{Reversibility \label{subrev}}

The stability of the integration of molecular dynamics equations of motion
in connection
with Hybrid Monte Carlo algorithms for lattice QCD has been investigated in the
literature \cite{Liu:1998fs,Frezzotti:1998eu,Joo:2000dh}. We report here on
our experiences looking at the reversibility of the integration of the equations
of motion with $\npf=2$ pseudo-fermion fields.
We compare how quantities which we monitor to check reversibility
behave when the lattice spacing or the trajectory length are changed.

\subsubsection{Diagnostics}

We perform reversibility tests by integrating over one trajectory of length $\tau$
``forward'' in molecular dynamics time, reversing at the end of the trajectory the
signs of the momenta
and integrating ``backward'' in time over one trajectory of
the same length $\tau$. We obtain a cycle:
\be
 \{\pi,U\} \stackrel{\displaystyle\tau}{\longrightarrow}
 \{\pi^{\prime},U^{\prime}\} \rightarrow \{-\pi^{\prime},U^{\prime}\}
 \stackrel{\displaystyle\tau}{\longrightarrow} \{\pi^{\pp},U^{\pp}\}
\ee
Reversibility means that the final configuration $\{\pi^{\pp},U^{\pp}\}$
is equal to the starting one $\{\pi,U\}$. We quantify irreversibility
by defining the quantities
\bea
 \DHc  & = & \left| H(\pi,U,\phi_i) - H(\pi^{\pp},U^{\pp},\phi_i) \right| 
 \label{dhcycle} \\
 \DU^2 & = & \frac{1}{36L^4}\sum_{x,\mu,c,c^{\prime}} 
 \left| U(x,\mu)_{c\,c^{\prime}} - U^{\pp}(x,\mu)_{c\,c^{\prime}} \right|^2 
 \label{du} \,,
\eea
where $c,c^{\prime}$ are the color indices. The step-size is chosen 
as in our production runs to keep the
acceptance close to 80\% with $\tau=1$ and the relative residue is set to
$\eps^2=10^{-13}$.
\begin{figure}[t]
 \begin{center}
     \includegraphics[width=11cm]{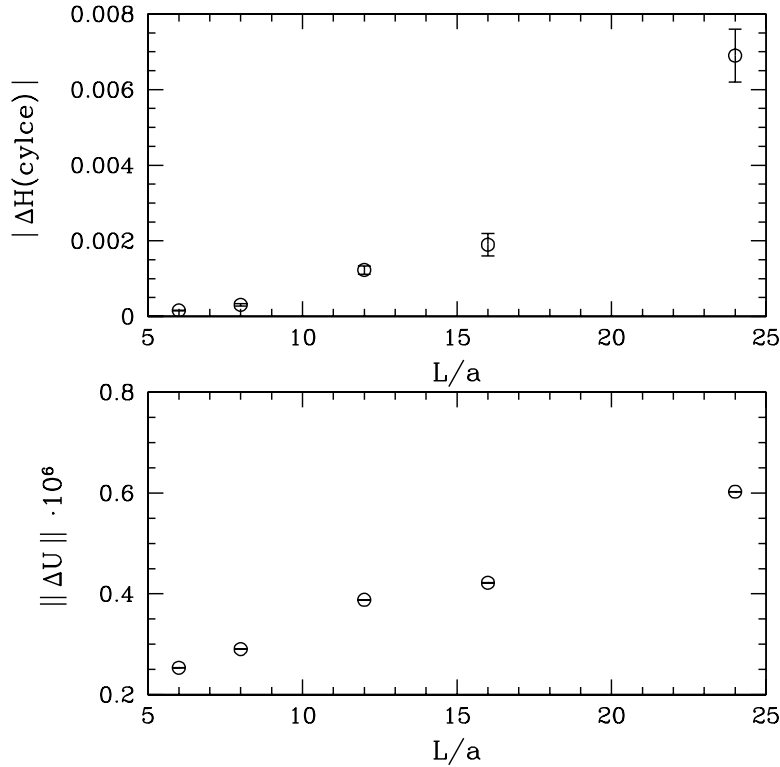}
 \end{center}
 \vspace{-0.5cm}
 \caption{Reversibility violations as functions of the lattice spacing for $\npf=2$.
The renormalized coupling is 2.5. The trajectories used to compute the quantities
in \eq{dhcycle} and \eq{du} have length $\tau=1$.}
 \label{f_revvol}
\end{figure}
\begin{figure}[t]
 \begin{center}
     \includegraphics[width=11cm]{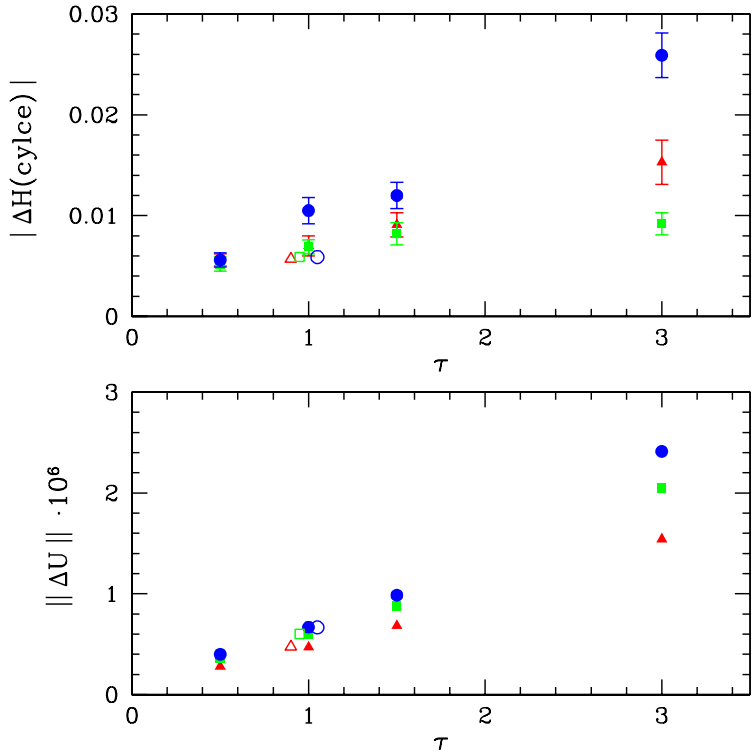}
 \end{center}
 \vspace{-0.5cm}
 \caption{Reversibility violations as functions of the trajectory length on
$L/a=24$ lattices for $\npf=2$. The renormalized couplings are 1.1 (triangles), 
2.5 (squares) and 5.7 (circles).
The open symbols for $\tau=1$, slightly displaced for clarity,
are obtained by switching off the fermions during the MD evolution.}
 \label{f_revtau}
\end{figure}

In \fig{f_revvol} we show the irreversibility as a function of the lattice spacing.
We plot averages of the quantities \eq{dhcycle} and \eq{du} computed
on several configurations, which are well spaced in the Monte Carlo history of our
production runs to avoid autocorrelations. The trajectories for these reversibility
tests have all length $\tau=1$, the step-size ranges from 1/5 on $L/a=6$ to
1/16 on $L/a=24$ lattices.
The renormalized coupling has the value 2.5,
so increasing $L/a$ decreases the lattice spacing by the same factor.

In \fig{f_revtau} we show the irreversibility as a function
of the trajectory length $\tau$ used to compute the quantities
\eq{dhcycle} and \eq{du}. 
Varying $\tau$ for fixed step-size gives us a handle to artificially
change the amount of reversibility violations.
We plot results for three different couplings 1.1 (triangles),
2.5 (squares) and 5.7 (circles) on $L/a=24$ lattices.
The step-sizes are 1/12, 1/16 and 1/18 respectively. 
The quantities in \eq{dhcycle} and \eq{du} are
averaged over 30 well spaced configurations taken from our production runs.
The relative increase of $\DHc$ and $\DU$ with increasing $\tau$ 
is of comparable magnitude.
In general the irreversibility is slightly larger for larger coupling.

For $\tau=1$ we also considered
the case when the fermions are switched off in the integration of
the equations of motion.
This corresponds to the open symbols in \fig{f_revtau}, which are slightly displaced
horizontally to distinguish them. We conclude that the
reversibility violations coming from the gauge part of the force which governs
the evolution of the momenta are not negligible with respect to the ones coming
from the pseudo-fermion part. 

\subsubsection{Influence on observables}

To check that we can safely simulate a trajectory of length $\tau=1$ with
the relative residue in the CG iterations set to $\eps^2=10^{-13}$,
we repeated one simulation on $L/a=24$ lattices at coupling 1.1
with trajectory length $\tau=0.5$, keeping the other parameters unchanged.
As it is shown in \fig{f_revtau} the reversibility violations for $\tau=0.5$ are
smaller than for $\tau=1$ and so we consider simulations with
$\tau=0.5$ as a safe reference for the correct mean values of observables
we are interested in.

The results are shown in \tab{t_tau} and confirm our expectation.
In addition we do not see a difference in efficiency between $\tau=0.5$ and $\tau=1$
within our errors on $\tauint(\ZP)$.
 \begin{table}[ht] 
   \centering
   \begin{tabular}{|ccccccc|}
    \hline
    $\tau$ & $\eps^2$ & $\ZP$ & $\tauint(\ZP)\;[\MD]$ & plaq. & $am$ & $\Iam1$ \\[0.5ex]
  \hline\hline
1 & $10^{-13}$ & 0.7828(13) & 26(8) & 0.78876(1) & $-0.00076(2)$ & 1.012(10) \\
0.5 & $10^{-13}$ & 0.7813(13)& 24(8) & 0.78876(1) & $-0.00074(1)$ & 0.998(7) \\
  \hline 
 \end{tabular} 
\caption{Comparison of simulation results for $\npf=2$ using trajectory length $\tau=1$ and
$\tau=0.5$. Results stem from $L/a=24$ lattices at coupling 1.1.}
\label{t_tau}
\end{table}

From these results and from the ones in \sect{subpre}
we conclude that it is safe for our purposes to simulate with 
the relative residue set to $\eps^2=10^{-13}$ in the CG inversions
and with trajectory length $\tau=1$.

\subsection{Performance}

In this section we compare the performance of our simulations using $\npf=2$
pseudo-fermion fields with simulations using $\npf=1$.
\begin{table}[h] 
 \centering
 \begin{tabular}{|cccccc|}
  \hline
  $\npf$ & $\delta\tau$  & \# traj. & $\Pa$ & plaq. & $\Iam1$ \\[0.5ex]
  \hline\hline
2 & 1/18 & 2744 & 83\% & 0.62725(1) & 1.020(9) \\
1 & 1/36 & 160 & 49\% & 0.62726(5) & 0.81(12) \\
1 & 1/40 & 168 & 78\% & 0.62731(5) & 0.96(4) \\
1 & 1/50 & 168 & 89\% & 0.62732(3) & 1.036(27) \\
  \hline 
 \end{tabular} 
\caption{Comparison of the step-size between $\npf=1$ and $\npf=2$. Results
on $L/a=24$ lattices at coupling 5.7.}
\label{t_dtau}
\end{table}

First we look at the step-size which is required to get the same acceptance.
The acceptances in our simulations typically scatter around 80\%.
Following \cite{Gupta:1990ka,Takaishi:1999bi} we assume that
the acceptance probability $\Pa$ depends on the step-size $\delta\tau$ of the
improved Sexton-Weingarten integration scheme like
\be\label{accorr}
 \ln\Pa \approx 1 - \Pa \propto (\delta\tau)^4 \,.
\ee
This formula takes into account what are expected to be the dominant discretization
errors of the integration. We use \eq{accorr} for a small correction 
to $\delta\tau$ corresponding to $\Pa=80\%$ precisely.
\tab{t_dtau} shows results for simulations on $L/a=24$ lattices at coupling 5.7.
The acceptance 80\% is achieved
with $\delta\tau=1/17$ for $\npf=2$ and $\delta\tau=1/41$ for $\npf=1$.
This is more than a factor two difference.

\begin{figure}[t]
 \begin{center}
     \includegraphics[width=12cm]{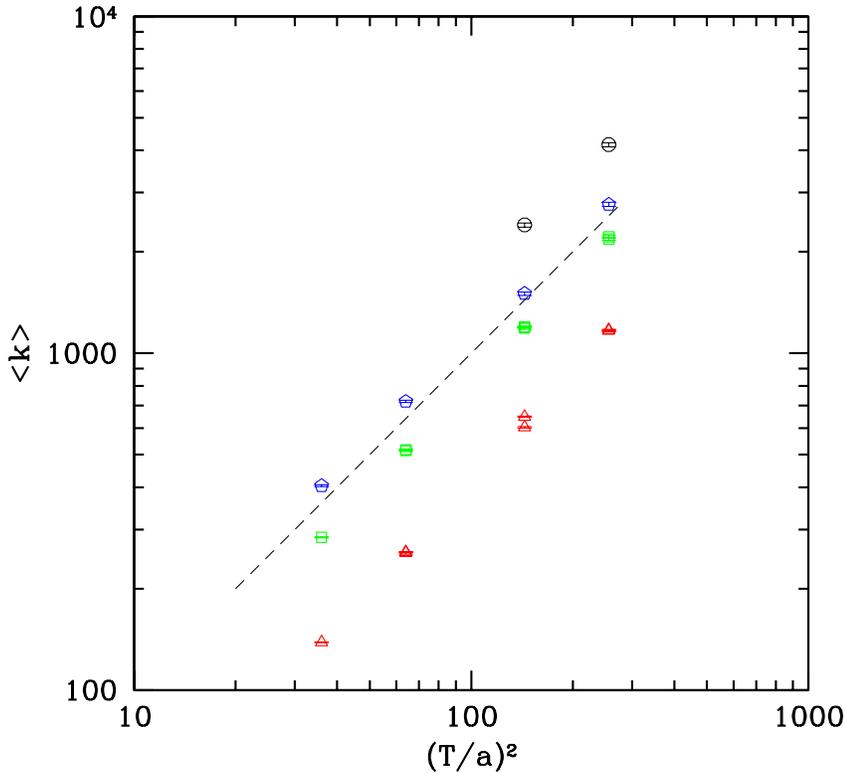}
  \end{center}
 \vspace{-1.5cm}
 \caption{Scaling of the average condition number $\ev{k}$ of $\hat{Q}^2$ 
with $(T/a)^2$.
Data for couplings 1.0 and 1.1 combined (triangles), 2.5 (squares), 3.3 (pentagons)
and 5.7 (circles).}
 \label{f_ksca}
\end{figure}
\begin{figure}[t]
 \begin{center}
     \includegraphics[width=12cm]{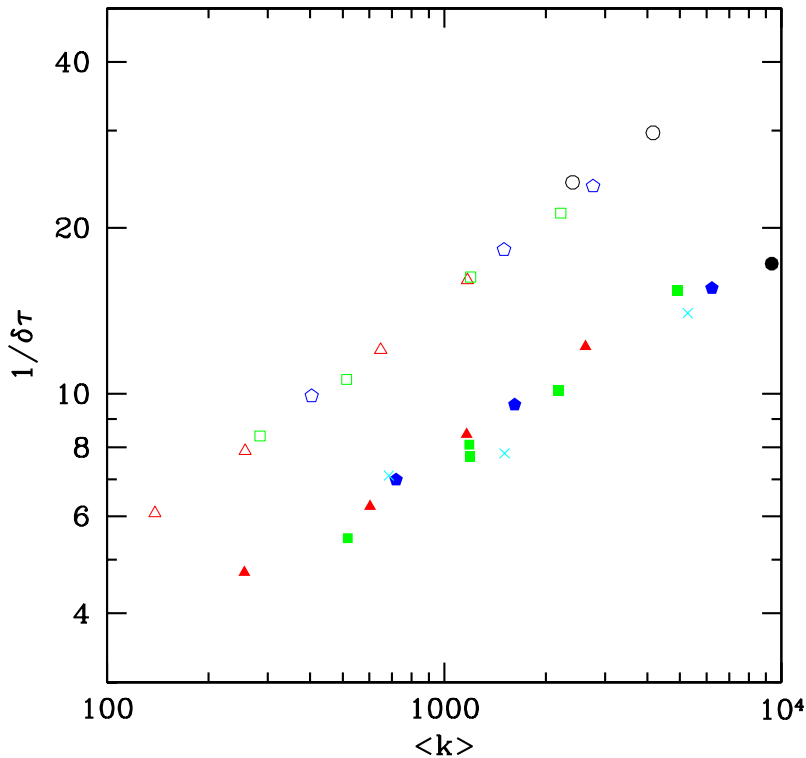}
  \end{center}
 \vspace{-1.5cm}
 \caption{Scaling of the inverse step-size $1/\delta\tau$ for 80\% acceptance
with the average condition number $\ev{k}$ of $\hat{Q}^2$.
Open symbols are for $\npf=1$, filled symbols and crosses for $\npf=2$
as explained in the text.}
 \label{f_dtausca}
\end{figure}
\begin{figure}[t]
 \begin{center}
     \includegraphics[width=12cm]{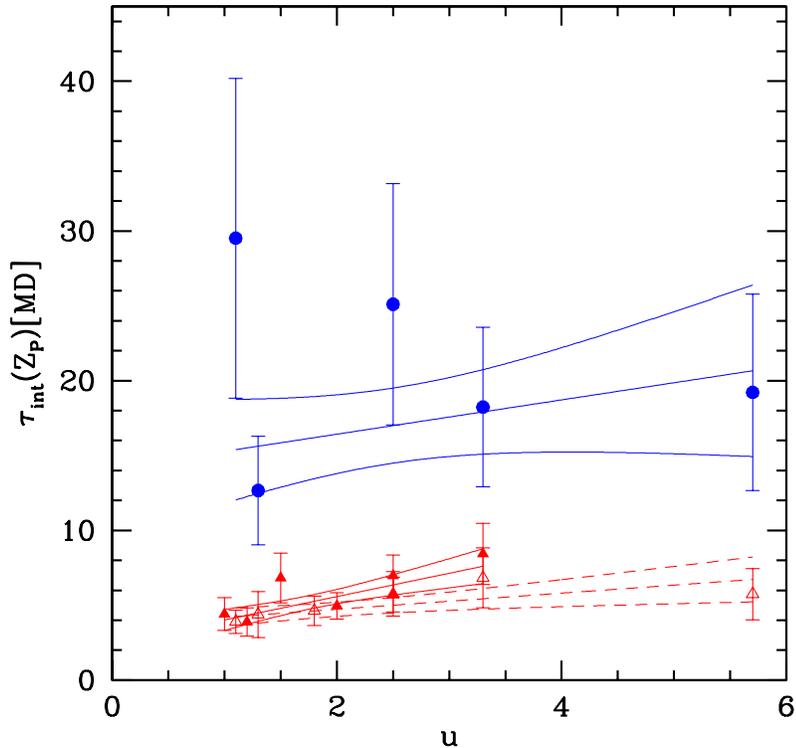}
  \end{center}
 \vspace{-1.5cm}
 \caption{The integrated autocorrelation time $\tauint(\ZP)\;[\MD]$ as function
of the renormalized coupling $u$. Data for $L/a=12$ with $\npf=1$ (open
triangles), $L/a=12$ with $\npf=2$ (filled triangles) and $L/a=24$ with $\npf=2$ 
(filled circles) are shown together with linear fits and their 1$\sigma$ error
bands.}
 \label{f_tauintmeas}
\end{figure}
\begin{figure}[t]
 \begin{center}
     \includegraphics[width=12cm]{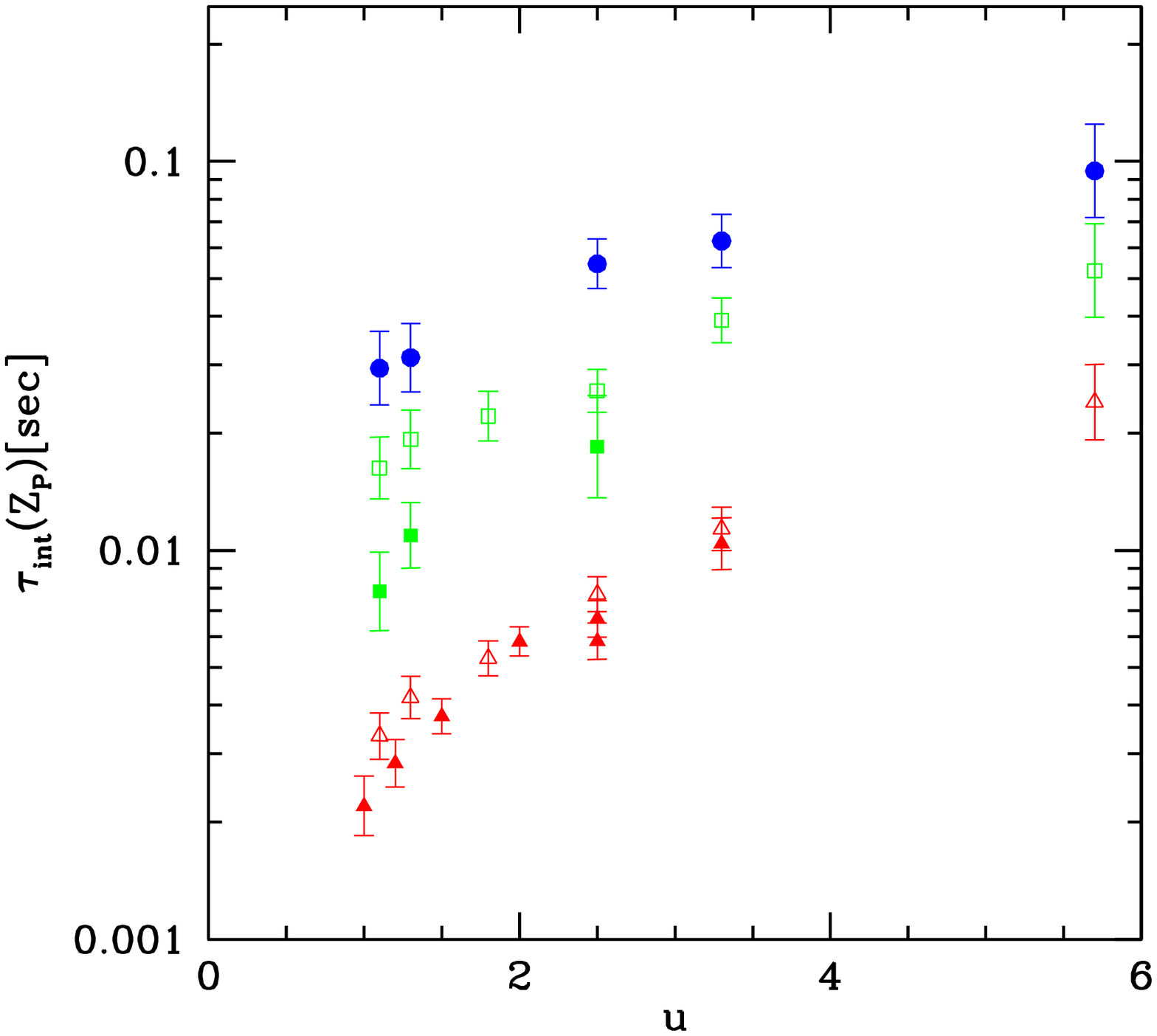}
  \end{center}
 \vspace{-1.5cm}
 \caption{The integrated autocorrelation time $\tauint(\ZP)\;[\sec]$ as function
of the renormalized coupling $u$. Data for lattices $L/a=12$ (triangles), $L/a=16$ (squares)
and $L/a=24$ (circles) are shown, open symbols are for $\npf=1$ and
filled symbols for $\npf=2$. These data are obtained from \eq{tauint} taking the
linear fits for $\tauint(\ZP)\;[\MD]$.}
 \label{f_tauintsec}
\end{figure}
In \fig{f_ksca} we plot our available data for the average condition number
$\ev{k}$ of the operator $\hat{Q}^2$ as function of $(T/a)^2$.
Data are for couplings 1.0 and 1.1 combined (triangles), 2.5 (squares), 3.3 (pentagons)
and 5.7 (circles). They confirm well the relation \eq{cnev} and we use it to extrapolate
$\ev{k}$ where we do not have measurements of the smallest and highest eigenvalue of
$\hat{Q}^2$. To guide the eye we draw the dashed line $\ev{k} = 10(T/a)^2$. 

In \fig{f_dtausca} the inverse step-size 
(i.e. the number of steps in one trajectory with $\tau=1$)
is plotted against $\ev{k}$ in a doubly logarithmic plot. The step-sizes have been
corrected according to \eq{accorr} to normalize the acceptance to 0.80.
Data are for couplings 1.0 and 1.1 combined (triangles), 2.5 (squares), 3.3 (pentagons) and
5.7 (circles). Open symbols are for $\npf=1$ and filled symbols for $\npf=2$. 
In addition we plot data for $\beta=5.2$, $\kappa=0.1355$,
$L/a=6,\,8$ and 12 runs (crosses) with $\npf=2$.
The points both for $\npf=1$ and $\npf=2$ fall on almost parallel 
straight lines showing remarkable scaling for the different couplings used.
A parallel displacement in a doubly logarithmic plot corresponds to a change
in the pre-factor of the power law relating $1/\delta\tau$ to $\ev{k}$. 

In \fig{f_tauintmeas} we plot the integrated autocorrelation time
$\tauint(\ZP)\;[\MD]$ as function of the renormalized coupling $u$.
A careful analysis of the error of $\ZP$ and $\tauint(\ZP)$
has been done using the method described in \cite{Wolff:2003sm}.
Data are for $L/a=12$ (triangles) and $L/a=24$ (circles) lattices, open symbols refer to
simulations with $\npf=1$ and filled symbols to $\npf=2$.
In the coupling range simulated there is no significant dependence on the coupling.
We show in the plot the results with error bands of linear fits to the data.
Similar fits have been done for the $L/a=16$ lattices not shown in \fig{f_tauintmeas}.
The fits are used to estimate the error of $\ZP$, so their uncertainty
influence the error of this error.
We do not observe a significant difference between $\npf=1$ and $\npf=2$.
This supports the results of \cite{Hasenbusch:2002ai} and the
assumption made in \cite{AliKhan:2003br}.

To see the difference between $\npf=1$ and $\npf=2$ we plot in \fig{f_tauintsec}
the autocorrelation time $\tauint(\ZP)\;[\sec]$ as function
of the renormalized coupling $u$. In these units
$\tauint(\ZP)$ becomes a direct measure of the CPU cost of the algorithms. Data are for
$L/a=12$ (triangles), $L/a=16$ (squares) and $L/a=24$ (circles) lattices,
open symbols are for $\npf=1$ and filled symbols for $\npf=2$.
The data shown are obtained from \eq{tauint} taking the values and errors of
$\tauint(\ZP)\;[\MD]$, which result from the linear fits in $u$.
A systematic difference between $\npf=1$ and $\npf=2$ is seen for the
$L/a=16$ lattices, the smallest couplings show a reduction of
$\tauint(\ZP)\;[\sec]$ by about a factor two for $\npf=2$.
We did not run with $\npf=1$ on $L/a=24$ lattices because, according to our
expectations based on the results for the $L/a=16$ lattices, this would be demanding
too much CPU time. In \fig{f_tauintsec} we see that the value of
$\tauint(\ZP)\;[\sec]$ for the $L/a=16$ simulation with $\npf=1$ at the largest
coupling 5.7 is within error almost compatible with the value for the $L/a=24$
simulation with $\npf=2$ at the same coupling.
The gain in CPU cost can be larger than a factor two for the $L/a=24$ simulations
when using $\npf=2$ instead of $\npf=1$, as supported by the results for the
step-size in \tab{t_dtau}.

Based on the step-sizes shown in \fig{f_dtausca} and the results for
$\tauint(\ZP)\;[\MD]$ shown in \fig{f_tauintmeas} we would expect a
gain of a factor two in CPU cost for Hybrid Monte Carlo using $\npf=2$
pseudo-fermion fields compared to $\npf=1$.
There is a computational overhead for using $\npf=2$ instead of $\npf=1$
pseudo-fermions, which comes from the inversion of $\hat{Q}^2+\rho^2$
in addition to the inversion of $\hat{Q}^2$, see \eq{Xvec}.
This overhead compared to the number of CG iterations required for the inversion of
$\hat{Q}^2$ ranges from 20-30\% on $L/a=12$ lattices to 10-15$\%$ on $L/a=24$
lattices and diminishes as the coupling $u$ increases. This explains why
the results for the $L/a=12$ lattices shown in \fig{f_tauintsec}
indicate a smaller gain in CPU cost using $\npf=2$ than for larger lattices.

One should keep in mind that these results have been obtained on small
and intermediate physical volumes. The situation in larger volumes might be
even better for $\npf=2$, as our results for coupling 5.7 indicate.

%% file: sect4.tex
\section{Conclusions}

We investigated the Hybrid Monte Carlo algorithm using two pseudo-fermion fields
to simulate two massless flavors in the Schr\"odinger functional O($a$) improved
theory. This study is part of large scale simulations to compute the running
of the renormalized quark mass with the physical lattice size $L$ from the running
of the renormalization factor $\ZP$ of the pseudoscalar density.

Our results show a gain in CPU cost of a factor two when using two instead of
one pseudo-fermions. This gain comes from the larger step-size and is based
on the results for the integrated autocorrelation time of $\ZP$. Since our
lattices have $L\approx1\fm$ at most, it might well be that the gain is larger
on larger volumes \cite{MHtsukuba}. 

Our simulations are performed with single precision arithmetics. We tested
the effects of numerical precision and reversibility of the integration of
the molecular dynamics equations of motion on observables. We confirm that
our simulations satisfy the requirements imposed both on numerical precision
and reversibility.
\\[0.5cm]
{\bf Acknowledgements.} We are grateful to M. Hasenbusch for discussions and
to M. L{\"u}scher for feedback on the manuscript.
We thank NIC/DESY Zeuthen for allocating computer time on the APEmille machine
indispensable to this project and the APE group for their professional and
constant support.
This work is supported in part by the European Community's Human Potential
Programme, contract HPRN-CT-2000-00145, ``Hadron Phenomenology from Lattice QCD''
and by the Deutsche Forschungsgemeinschaft in the SFB/TR 09-03,
``Computational Particle Physics''.

%% file: appa.tex
\section{Simulation parameters \label{simpar}}

\begin{table}[h] 
 \centering
  \begin{tabular}{lll||lll}
   \hline
   $L/a$ & $\beta$ & $\kappa$ & $L/a$ & $\beta$ & $\kappa$ \\
   \hline
 6  & 9.5000   & 0.1315322 &  6  & 6.6085  & 0.1352600 \\
 8  & 9.7341   & 0.1313050 &  8  & 6.8217  & 0.1348910 \\
12  & 10.05755 & 0.1310691 & 12  & 7.09300 & 0.1344320 \\[2.0ex]

 6  & 8.5000   & 0.1325094 &  6  & 6.1330  & 0.1361100 \\
 8  & 8.7223   & 0.1322907 &  8  & 6.3229  & 0.1357673 \\
12  & 8.99366  & 0.1319754 & 12  & 6.63164 & 0.1352270 \\[2.0ex]
                                
 6  & 7.5000   & 0.1338150 &  6  & 5.6215  & 0.1366650 \\
 8  & 7.7206   & 0.1334970 &  8  & 5.8097  & 0.1366077 \\
12  & 8.02599  & 0.1330633 & 12  & 6.11816 & 0.1361387 \\
   \hline
  \end{tabular} 
 \caption{Summary of simulation parameters.}
 \label{t_simpar}
\end{table}
\noindent
In \tab{t_simpar} we list the lattice sizes and bare parameters of our simulations.
These are divided in six groups of three sets each.
In each group $\beta$ and $\kappa$ have been tuned for the lattice sizes $L/a=6,8,12$
to keep the renormalized coupling $\gbar^2(L)$ at an approximately  
fixed value $u$
(see Ref. \cite{Bode:2001jv} for a precise definition of $\gbar^2(L)$),
in the left column $u=1.0,1.2,1.5$, in the right column $u=2.0,2.5,3.3$ in
the order from top to bottom.
For each bare parameter set we also simulated the doubled lattice sizes $2L/a$.
The renormalized coupling then takes the values $u=1.1,1.3,1.8$
and $u=2.5,3.3,5.7$ in the same ordering as before.

The O($a$) boundary improvement coefficient of the Wilson plaquette gauge action
$c_{\rm t}$ is set to the perturbative value \cite{Bode:1999sm}
\bea
 c_{\rm t} & = & 1-0.050718g_0^2-0.030g_0^4 \,. \label{ct}
\eea
For the left part of \tab{t_simpar} except in the simulations with
$\beta=7.7206$ and $\kappa_c=0.1334970$ we take the 1-loop expression,
all the other simulations use the 2-loop expression of $c_{\rm t}$.

As concerns the O($a$) improvement of the Wilson-Dirac operator,
the clover coefficient $c_{\rm sw}$ is computed using the non-perturbative
fit-formula given in \cite{Jansen:1998mx}
\bea
 c_{\rm sw} & = & \frac{\displaystyle 1-0.454g_0^2-0.175g_0^4+0.012g_0^6+0.045g_0^8}
                       {\displaystyle 1-0.720g_0^2} \,.
\eea
The coefficient $\tilde{c}_{\rm t}$ is set to the
1-loop perturbative value \cite{Sint:1997jx}
\bea
 \tilde{c}_{\rm t} & = & 1-0.01795g_0^2 \,.
\eea

Finally we use for the O($a$) improvement coefficient of the axial current
$c_{\rm A}$ the 1-loop perturbative value \cite{Sint:1997jx}
\bea
 c_{\rm A} & = & -0.007573g_0^2 \,.
\eea